\documentclass[fleqn,12pt]{wlscirep}
\usepackage{wrapfig, framed, caption}
\usepackage[utf8]{inputenc}
\usepackage{heuristica}
\usepackage{graphicx}
\usepackage{wrapfig, framed, caption}
\usepackage{tcolorbox}
\usepackage{xr}
\usepackage{hyperref}
\usepackage{xcolor}

\makeatletter
\newcommand*{\addFileDependency}[1]{
  \typeout{(#1)}
  \@addtofilelist{#1}
  \IfFileExists{#1}{}{\typeout{No file #1.}}
}
\makeatother

\usepackage{setspace}
\usepackage[normalem]{ulem}
\useunder{\uline}{\ulined}{}
\usepackage{xparse}
\newsavebox{\fminipagebox}
\NewDocumentEnvironment{fminipage}{m O{\fboxsep}}
 {\par\kern#2\noindent\begin{lrbox}{\fminipagebox}
  \begin{minipage}{#1}\ignorespaces
 \end{minipage}\end{lrbox}%
  \makebox[#1]{%
    \kern\dimexpr-\fboxsep-\fboxrule\relax
    \fbox{\usebox{\fminipagebox}}%
    \kern\dimexpr-\fboxsep-\fboxrule\relax
  }\par\kern#2
 }

\title{Decoding the Manhattan Project’s Network: Unveiling Science, Collaboration, and Human Legacy
}

\author[*]{Mil\'an Janosov}

\affil[*]{milan@janosov.com, \href{www.janosov.com}{www.janosov.com}}

\begin{document}

\maketitle


\section*{Abstract}
{\small

The Manhattan Project was one of the largest scientific collaborations ever undertaken. It operated thanks to a complex social network of extraordinary minds and it became undoubtedly one of the most remarkable intellectual efforts of human history. It also had devastating consequences during and after the atomic bombings of Hiroshima and Nagasaki. Despite the loss of hundreds of thousands of human lives during the bombing and the subsequent events, the scientific journey itself stands as a testament to human achievement, as highlighted in Christopher Nolan’s film portrayal of Oppenheimer.

}

\vspace{0.5cm}
{\small {\bf Keywords}: network science, social network analysis, Manhattan project, data science}

\vspace{1.0cm}
{\it \hspace{-1cm} Published in Nightingale, Journal of the Data Visualization Society, September 12, 2023~\cite{nightingale}.  Edited by Kathryn Hurchla.}
\vspace{1.0cm}

The scientific literature on collaboration, particularly the role of network connections in achieving success, is robust and has been further enriched by the current data boom. This wealth of data, represented by for instance millions of scientific papers, is exemplified in works such as “The Science of Science,” by D. Wang and A. L. Barabási~\cite{sos}. Utilizing network analysis to uncover the intricate connections within the Manhattan Project aligns with my perspective as a physicist turned network scientist. Without further ado, here’s how I mapped the Manhattan Project into data and used that to create a network visualization of this historically significant collaborative project.

\section{Collecting Data}

As with many data science projects, the first question revolved around data selection. While scientific publication data might seem logical, given the project’s scientific nature, this approach proved inadequate. The main reason for this was two-fold: First, some of the most important documents and papers could still be classified; and also, not everyone was active in science, as the operation was also heavily intertwined with politics and the military. Thus, resorting to collective wisdom, my focus shifted to Wikipedia, a global crowdsourced encyclopedia and a potential data source. Wikipedia offers a list of notable personnel connected to the project~\cite{wiki}, encompassing more than 400 contributors from various fields. I used  a straightforward web-scraping technique to collect data from Wikipedia—a total of 452 usable profiles. Then I manually categorized each person based on occupation, leading to the distribution outlined in Table \ref{table:table1}.

\begin{table}[!hbt]
\centering
\begin{tabular}{cc}
\textbf{Occupation} & \textbf{Ratio} \\
Physicist           & 51.99\%        \\
Chemist             & 17.7\%         \\
Engineer            & 9.29\%         \\
Other               & 6.19\%         \\
Army officer        & 5.53\%         \\
Mathematician       & 3.54\%         \\
Biologist           & 1.99\%         \\
Spy                 & 1.55\%         \\
Physician           & 1.11\%         \\
Computer scientist  & 1.11\%        
\end{tabular}
\caption{Occupation Distribution of Notable Manhattan Project Contributors.}
\label{table:table1}
\end{table}

The list, not entirely surprisingly, is topped by physicists, followed by chemists and engineers. However, exploring the realm of science, particularly those at the forefront of the Project, awaits. Let’s take the stories from the “Other” category. This group collects contributors’ primary occupations that appeared infrequently and seemed unrelated to a scientific project focused on weaponry development. Among these unconventional contributors are Wolfrid Rudyerd Boulton, an American ornithologist, who also happened to become responsible for monitoring the supply of uranium ore from the Belgian Congo, and Edith Warner, a tea room owner in Los Alamos whose role was said to have profoundly impacted researchers’ morale.

Some other notable “other” figures include Charlotte Serber, a journalist, statistician, librarian, and the sole female laboratory group leader in Los Alamos. Ben Porter defies categorization, too, embracing roles as an artist, writer, publisher, performer, and physicist—later exhibiting work at New York’s Museum of Modern Art. The selection concludes with James Edward Westcott, a notable Manhattan Project photographer, and Donald Lindley Harvey, a professional basketball player turned Army member contributing to the project.

\section{Constructing the Network}

\begin{figure}[!hbt]
\centering
\includegraphics[width=0.80\textwidth]{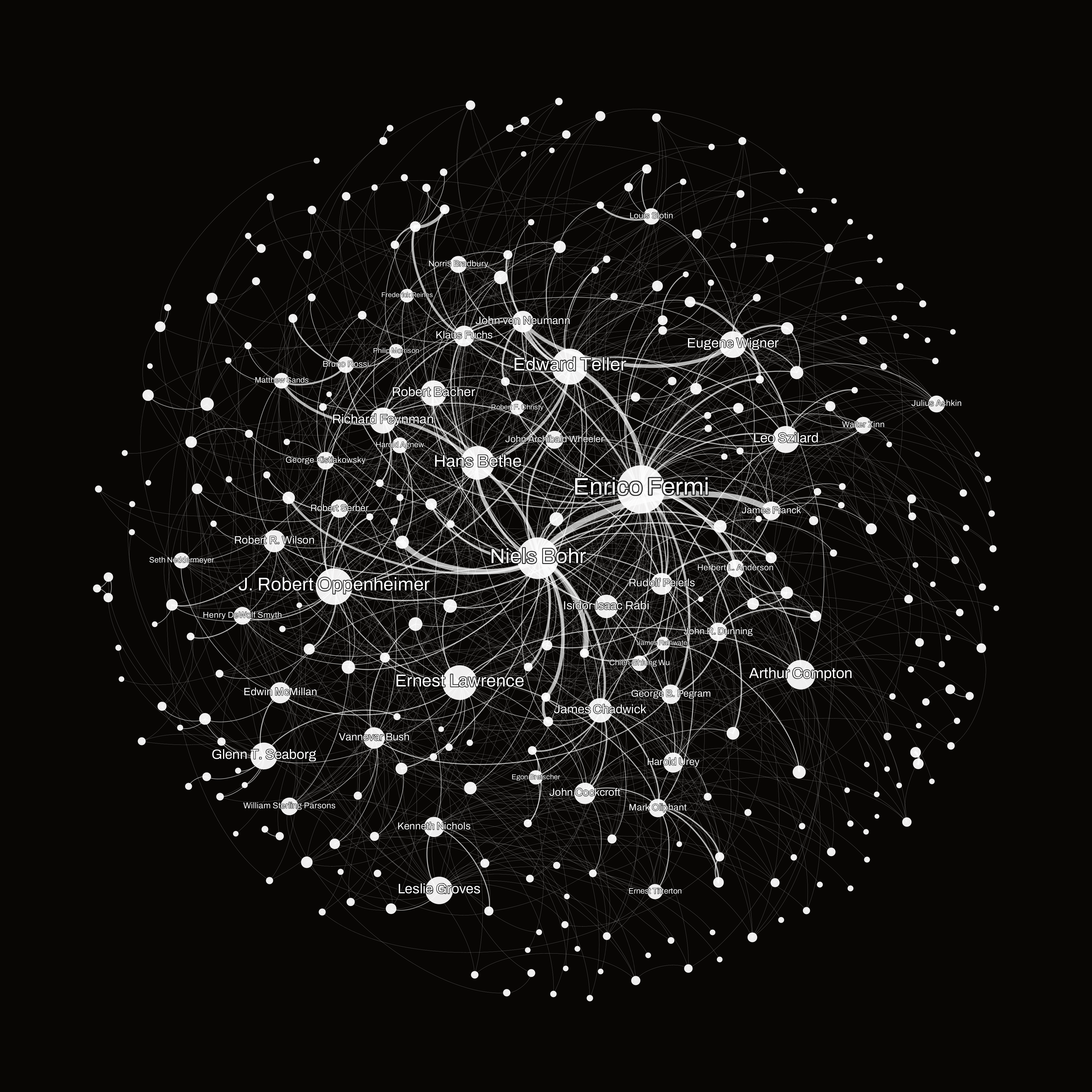}
\caption{The collaboration network behind the Manhattan Project. Each node represents a contributor, where two nodes are linked if their Wikipedia pages reference each other. The top 50 nodes with the largest number of connections are labeled.}
\label{fig:fig1}
\end{figure}

With the data in hand, I picked network science~\cite{netsci}, the science of connections that is perfect for elegantly deciphering complex structures such as the Manhattan Project’s collaboration patterns. Each network comprises nodes (entities) and links (references) that weave the intricate social fabric of the collaborating people. In this context, each node symbolizes a Manhattan Project contributor, with links forming between individuals whose Wikipedia pages reference one another. The number of shared references determines the link’s strength. Employing this straightforward framework, I arrived at a network of 316 individuals connected by 1,099 ties of various strengths.

\section{Infusing Color into Insight}

\begin{figure}[!hbt]
\centering
\includegraphics[width=0.80\textwidth]{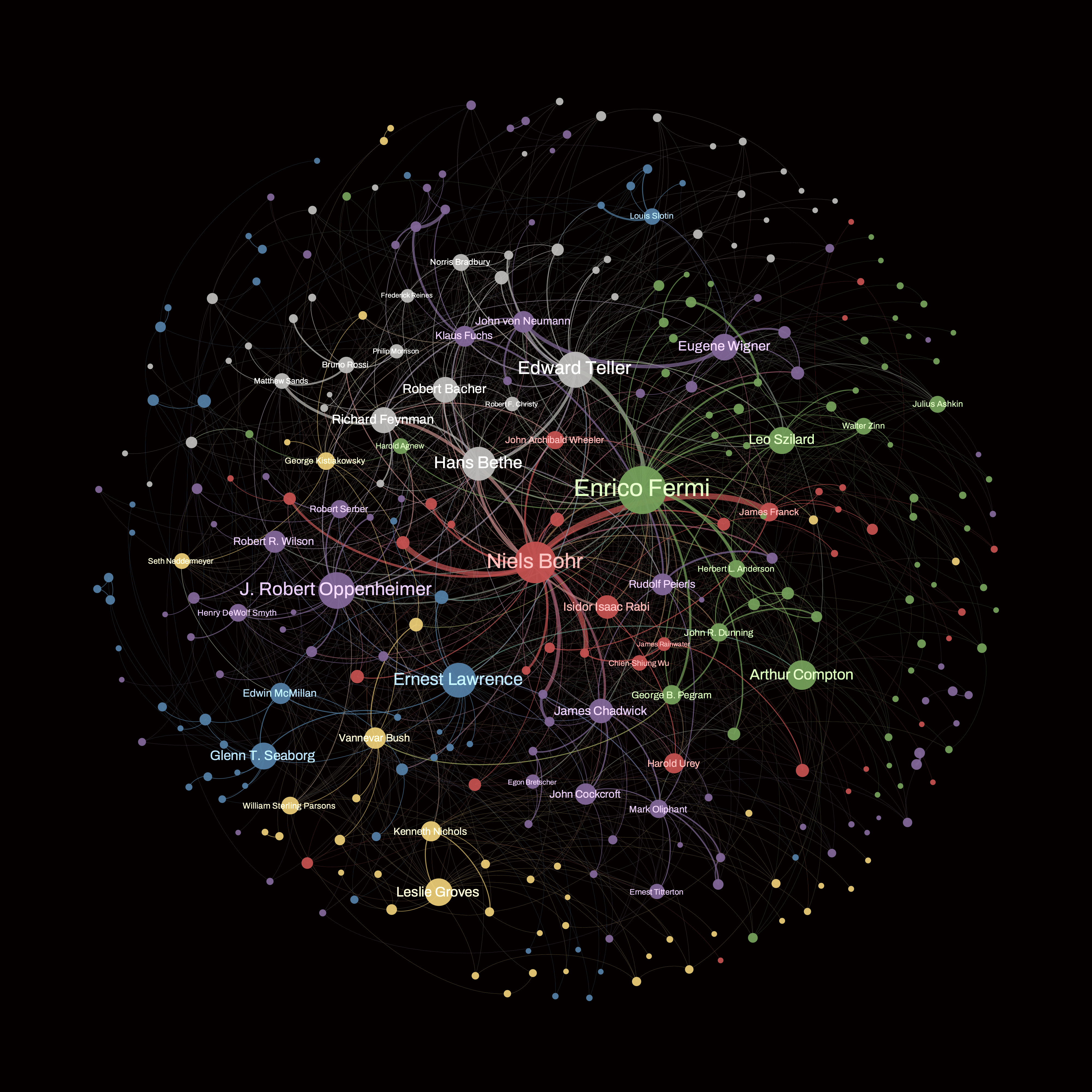}
\caption{The collaboration network behind the Manhattan Project shown in Figure \ref{fig:fig1}, where each node is colored based on the network community it belongs to.}
\label{fig:fig2}
\end{figure}

The next phase enriches the network visualization by introducing color—each hue representing a distinct network community or cluster. Defining these communities hinges on the methodology, but the general premise remains: Communities consist of nodes with a higher density of internal links than external ones~\cite{witcher, blondel2008fast}. In other words, nodes mostly linked to each other—as opposed to the rest of the network—belong to one community. The resulting visual, presented in Figure \ref{fig:fig2}, uncovers how contributors organize into closely connected clusters within the expansive Manhattan Project. In this Figure, each color encodes different communities.

\section{Deciphering the Network's Narrative}

With the vibrant visualization in Figure \ref{fig:fig2}, we are ready to read the collaboration network. Key players in modern physics pop out immediately, including Nobel laureates Arthur Compton, Enrico Fermi, Niels Bohr, and Ernest Lawrence, alongside geniuses like J. Robert Oppenheimer and Edward Teller. Yet, there is much more to the story and the patterns behind the connections than just a handful of hubs.

\begin{figure}[!hbt]
\centering
\includegraphics[width=0.80\textwidth]{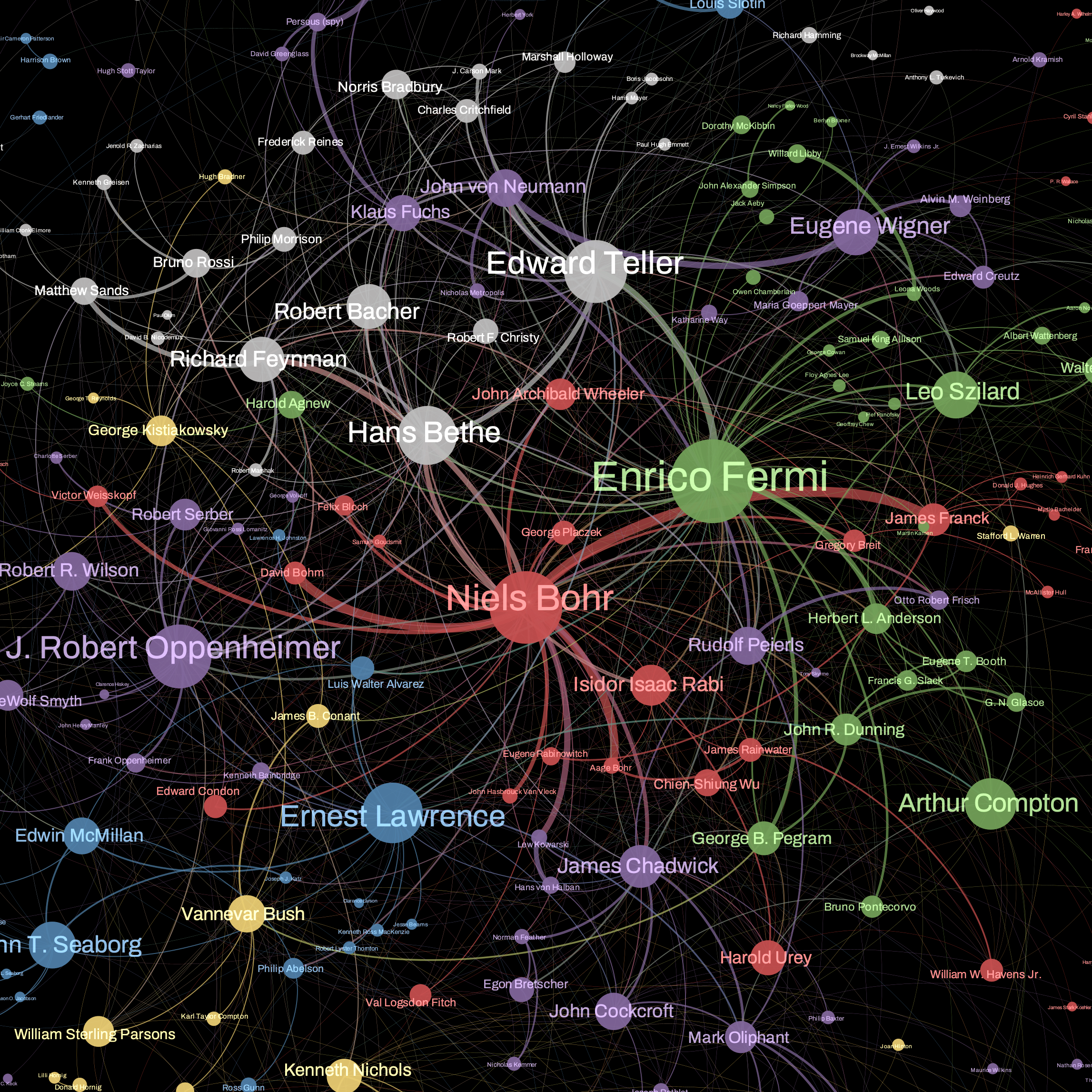}
\caption{A close-up of the collaboration network behind the Manhattan Project colored by network communities shown in Figure \ref{fig:fig2}, where each node is labeled.}
\label{fig:fig3}
\end{figure}

At the core of this network diagram lies the red community centered by the legendary Niels Bohr. Here, Bohr’s connections reveal his instrumental role in supporting refugee scientists during World War II, who also joined the Project, including people like Felix Bloch, James Franck, and George Placzek, all marked by red. Adjacent to Bohr’s realm resides a green cluster, highlighted by the Italian physicist Enrico Fermi. Fermi, together with his collaborators like Anderson, Szilárd, Compton, and Zinn, reached the milestone of the self-sustaining chain reaction using uranium and gave birth to the first nuclear reactor, the Chicago Pile-1.

While Eugene Wigner was most famous for his contribution to Chicago Pile-1, his links tie him closer to the purple community that seems to be scattered around the network. Wigner can be seen prominently in the upper-right corner of the network. This more decentralized community, having no one else but Oppenheimer as its key figure, also links the famous Mathematician John von Neumann, with purple, in the top-center part of Figure \ref{fig:fig3}, who. (He, along with Wigner was unfortunately left out of the blockbuster movie by Nolan.) With purple, we see several other leading scientists, such as James Chadwick in the bottom-center, who led the British team on the Project; Robert Wilson right next to Oppenheimer, who became the head of its Cyclotron Group; and the American physicist Robert Serber directly above Oppenheimer, who created the code names for all three design projects and the bomb, such as “Little Boy” and “Fat Man.” Finally, a few words about the gray cluster, which turned out to be the Theoretical Division, with stars like Edward Teller in the center, and Nobel laureates Richard Feynman (my personal favorite scientist) in the top left, and Hans Bethe in the center.

\begin{figure}[!hbt]
\centering
\includegraphics[width=0.80\textwidth]{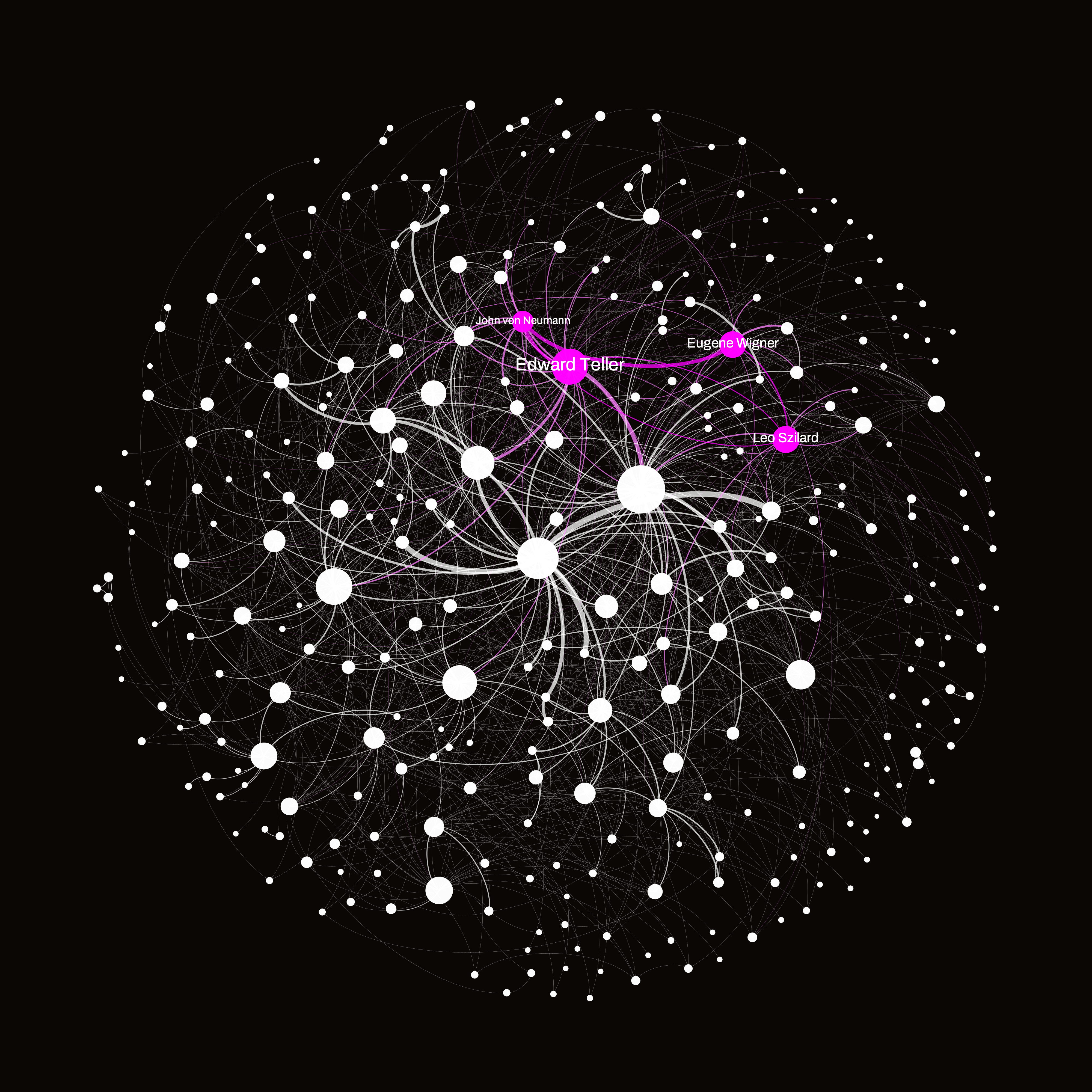}
\caption{A variant of Figure \ref{fig:fig2} highlighting the Martians – Edward Teller, Eugene Wigner, Leo Szilard, and John von Neumann.}
\label{fig:fig4}
\end{figure}

One last observation to a personal accord: At first sight, the connections between the Hungarian immigrant Martians~\cite{wiki2} Teller, Wigner, Szilard, and Neuman were hard to spot, despite their foundational role in the dawn of the atomic era and countless joint projects. However, once I highlighted them on the network, my expectations were quickly confirmed. They are all closely linked though not exclusively, meaning that they were also very well embedded in the American scientific community at that time (Figure \ref{fig:fig4}). This is probably best illustrated by the so-called Einstein Szilard letter, written by Szilard who also consulted with Teller and Wigner, and which was ultimately signed by and sent to President Roosevelt by Einstein. A fun fact about this letter: during those days, Einstein was spending his vacation on the beach, so Szilard visited him right there. And as Szilard didn’t own a driver’s license, Teller was driving him~\cite{marx}.

\section{Closing}

Beyond the pages of history, the project embodies the convergence of human endeavor—distinguished minds across varied disciplines united for a common goal. This analysis sheds some light on the complex patterns of collaboration and joint efforts that allowed such great minds to connect, work in teams, and succeed at such an enormous scale. Additionally, the way I built this network illustrates how network science can be applied to nearly any social system, quantitatively capturing the invisible relations, and putting them into a quantitative context.

\section{Disclaimer}

Several parts of this text were upgraded by AI tools, namely, Grammarly and ChatGPT 3.5, while the whole text was initially drafted and later updated by the human author.


\end{document}